\documentclass[aps,prb,showpacs,preprintnumbers,twocolumn,amsmath,amssymb]{revtex4-1}
\usepackage{graphicx}%
\usepackage{color}
\usepackage{bm}
\usepackage{amsmath}
\usepackage{braket}

\newcommand{\comment}[1]{}
\usepackage[]{hyperref}
\hypersetup{
  colorlinks   = true, 
  urlcolor     = blue, 
  linkcolor    = black, 
  citecolor   = blue 
}

\begin{document}

\title{Non-degenerate valleys in the half-metallic ferromagnet Fe/WS$_2$}

\date{\today}
\author{Omar Messaoudi$^{1,2}$, Julen Iba\~{n}ez-Azpiroz$^{1}$, Hamid Bouzar$^{2}$ and Samir Lounis$^{1}$}

\address{
$^{1}$Peter Gr\"{u}nberg Institut and Institute for Advanced Simulation, Forschungszentrum J\"{u}lich \& JARA, D-52425 J\"{u}lich, Germany
}
\address{$^{2}$ Laboratoire de Physique et Chimie Quantique (LPCQ), Universit\'e Mouloud Mammeri Tizi-Ouzou, BP 17 RP, 15000 Tizi-Ouzou, Alg\'erie}


\begin{abstract}

We present a first principles investigation of the electronic properties of monolayer WS$_2$ coated with an overlayer of Fe. Our \textit{ab initio} calculations reveal that the system is a half-metallic ferromagnet with a gap of $\sim 1$ eV for the majority spin channel. Furthermore, the combined effect of time-reversal symmetry breaking due to the magnetic Fe overlayer and the large spin-orbit coupling induced by W gives rise to non-degenerate K and K$'$ valleys. 
This has a tremendous impact on the excited state properties induced by externally applied circularly polarized light. 
Our analysis demonstrates that the latter induces a singular hot spot structure of the transition probability around the K and K$'$ valleys for right and left circular polarization, respectively. 
We trace back the emergence of this remarkable effect to the strong momentum dependent spin-noncollinearity of the valence band involved. As a main consequence, a strong valley-selective magnetic circular dichroism is obtained, making this system a prime candidate for spintronics
and photonics applications.

\end{abstract}

\maketitle

\section{Introduction.} 
Thin films exfoliated from van der Waals (vdW) bounded layers constructing the bulk crystal \cite{Novoselov2005a,Xu2013} demonstrate remarkable electronic and optical properties \cite{Butler2013,Splendiani2010,Mak2016,Schmidt2015}, and define a class of materials to which two-dimensional (2D) transition metal dichalcogenides (TMDs) belong to. These consist of multiple chemical compositions of the form MX$_2$ \cite{Structures1973}, where M is the transition metal and X the chalcogene atom. The resulting crystals are highly tunable and thus offer a wide variety of applications\cite{Choi2017} including energy storage\cite{Wu2014} and biosensors \cite{Wang2015}. Interestingly, when taken to the limit of a monolayer, group-VI
TMDs emerge as the most promising contenders, with MoX$_2$ and WX$_2$ (X= S, Se) being the top configurations. Indeed, these two types of compounds have been shown to be good semi-conductors with a direct
band gap laying in the near-infrared-to-visible range of the spectrum \cite{Splendiani2010,Mak2010,Zeng2013} , a property that makes them excellent
candidates for electronics and photonics applications \cite{Wang2014,Islam2014,Mak2010,Mak2016}. Moreover, given their low-cost fabrication process\cite{Wang2012,Chhowalla2013,Liu2012}, they constitute a basic building block for new material engineering, either by
alloying, coating or impurity deposition\cite{Zhang2016,Wang2014,Zhong2017}.

\

Group-VI TMDs arrange in a honeycomb lattice structure with broken inversion symmetry. Owing to this feature, a valley degree of freedom arises in the form of a pseudo-spin \cite{Xu2014,Wu2013,Yao2008,Song2015,Mak2014} that describes the two inequivalent but energy-degenerate band edges at the corners of the hexagonal Brillouin zone, the so called K and K$'$ valleys\cite{Xiao2012,Yao2008}. Manipulation of this new degree of freedom forms the basis of valleytronics\cite{Cao2011}. New valley-dependent optical selection rules \cite{Wang2017} emerge from the valley magnetic moment, where circularly polarized light excites electron-hole pairs in a specific valley depending on its polarization \cite{Suzuki2014,Song2015,Xiao2012,Sugawara2015}. Noteworthily, recent studies \cite{Srivastava2015,Wu2013,Li2014} have shown that applying a magnetic field normal to the surface causes opposite energy shifts in the K and K$'$ valleys as a consequence of contributions arising from the valley magnetic moment. This mechanism is described as a valley Zeeman effect, which enables the control of the the valley polarization via a  static magnetic field plus time dependent circularly polarized light. 

\
In this paper, we present a first principles study of the electronic 
properties of the Fe/WS$_2$ heterostructure. 
The goal achieved by adding a magnetic overlayer to a monolayer TMD is to break
time-reversal symmetry, thus mimicking the effect of the static magnetic field
leading to the aforementioned valley Zeeman effect.
Our \textit{ab initio} calculations
reveal that the system is a half-metallic ferromagnet with a gap of $\sim 1$ eV for the majority spin channel, 
thus making it attractive for spintronics and photonics applications. 
On top of that, due to the combined effect of time-reversal and inversion symmetry breaking,
the latter being strongly enhanced by the large spin-orbit coupling (SOC) induced by W,
the K and K$'$ valleys become non-degenerate.
This leads to several remarkable effects: on one hand, the momentum-dependent
spin-polarization of the valence band becomes highly non-collinear and furthermore
strongly asymmetric in the neighborhood of the K and K$'$ high symmetry points. 
As a second and most important effect, valley-dependent physics is also present in the excited state properties of the system, which we probe after applying circularly polarized light.
Our analysis reveals that the transitions induced by left and right polarized light
are structured in high probability hot spot regions around the K and K$'$ points, respectively.
Furthermore, the maximum probability at the two inequivalent valleys differs by nearly $33\%$,
giving rise to a strong valley-selective magnetic circular 
dichroism in the absorbed light that should be experimentally observable 
at energies around $0.3$ eV.

\

\section{Computational details.} The electronic ground-state calculations were performed using density functional theory with a plane wave basis as implemented in the QUANTUM ESPRESSO\cite{QE} package. A plane wave cutoff Ec = 150 Ry was needed to ensure convergence. Exchange-correlation effects were treated with the PBE-GGA\cite{Perdew1996} pseudopotentials. SOC was included in our calculations by using a fully relativistic norm-conserving pseudo-potential \cite{DalCorso2005} for W. The slab method was used to model the system where a thick vacuum layer of 30 \AA\ was used to avoid the interactions between periodic repetitions. The converged ground-states were obtained using a $15\times15\times1$ Monkhorst-Pack\cite{Pack1977} \textbf{k}-mesh for the self-consistent calculations and a rectangular \textbf{k}-mesh of $51\times51\times1$ for the non self-consistent ones. The structure has been fully relaxed with a convergence criterion of $1.0 \boldsymbol{\cdot} 10^{-4}$ Ry $\boldsymbol{\cdot}$ \AA$^{-1}$ for the forces acting on each individual atom and $1.0 \boldsymbol{\cdot} 10^{-8}$ Ry for the total energy.

Following the approach of refs. \onlinecite{Ibanez-Azpiroz2012,Ibanez-Azpiroz2013}, the transition matrix elements induced by an external time-dependent electric field have been calculated making use of maximally localized Wannier functions \cite{Wang2006,Lopez2012}. This allowed us to interpolate the matrix elements to a dense $1000\times1000$ \textbf{k}-mesh in the irreducible Brillouin zone at a low computational cost.
\begin{figure}[t]
\centering

\includegraphics[width=0.5\textwidth]{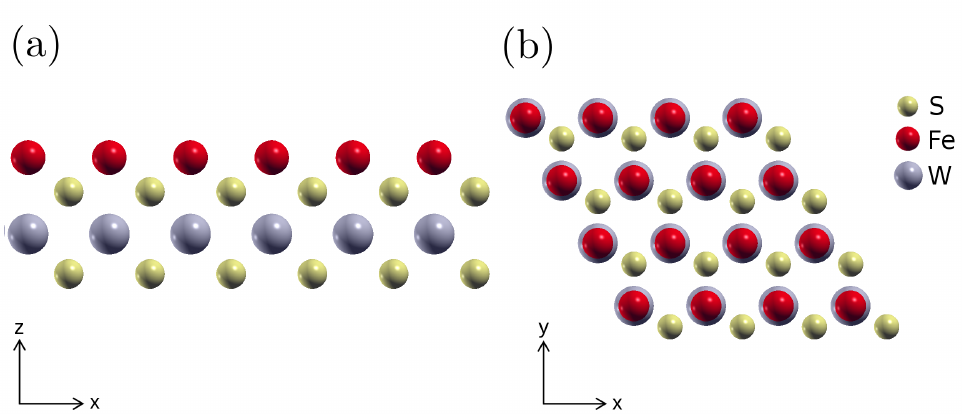}
\caption{(color online) Crystal structure of Fe/WS$_{2}$. (a) and (b) respectively show the side and top views.}
\label{fig:Crystal-Structure}

\end{figure}

\begin{figure}[t]
\centering

\includegraphics[width=0.5\textwidth]{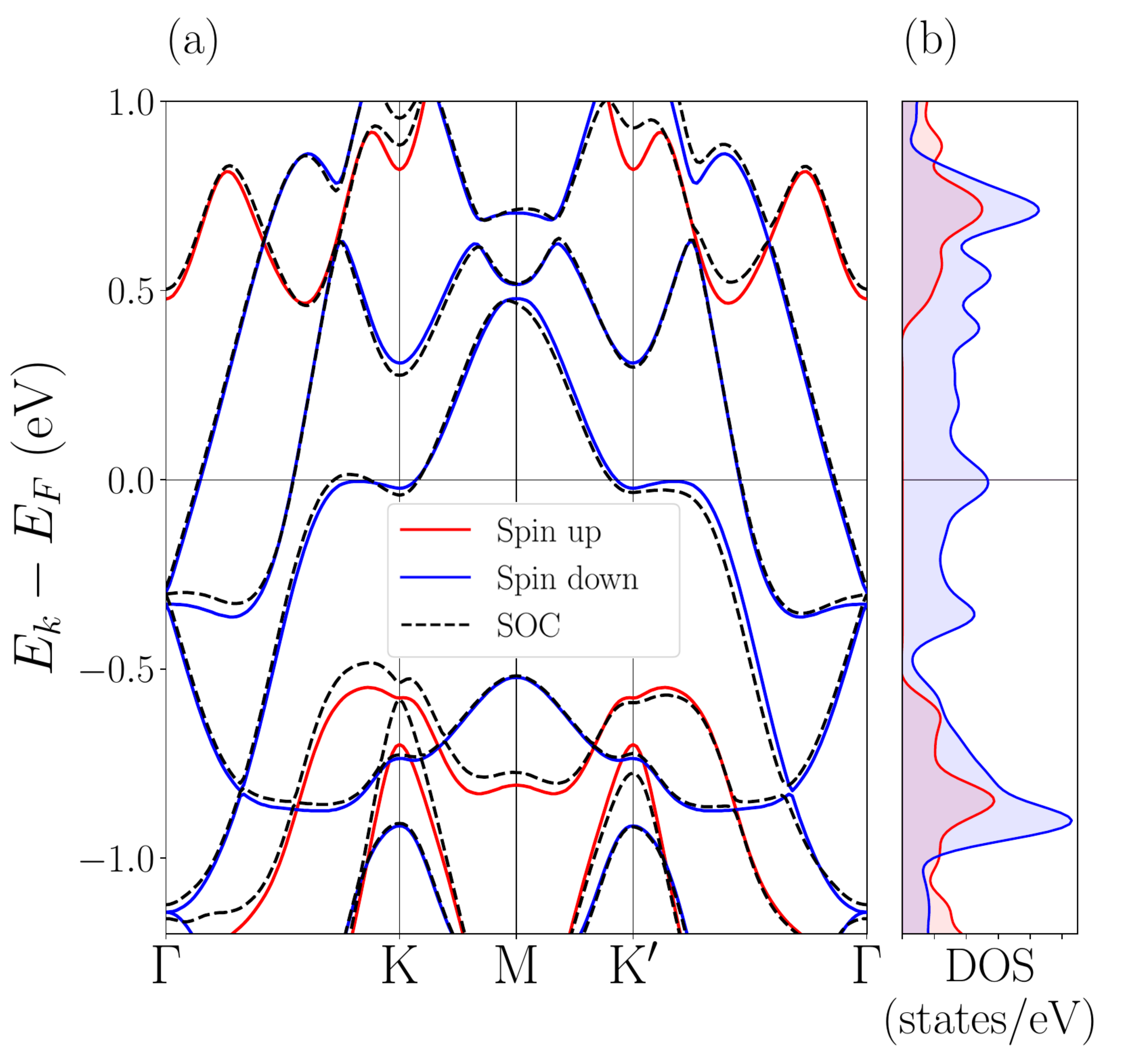}
\caption{(color online) Calculated electronic structure for  Fe/WS$_{2}$. (a) Calculated Band structure both in the scalar relativistic and fully relativistic cases. The red and blue lines represent respectively the spin up and spin down of the scalar relativistic calculation, while the dashed black line represents the fully relativistic bands. (b) Calculated scalar density of states, where as for the band structure the red and blue lines represent respectively the spin up and spin down contributions}
\label{fig:electron-structure}

\end{figure}

\begin{figure*}[t]
\centering
\includegraphics[width=1\textwidth]{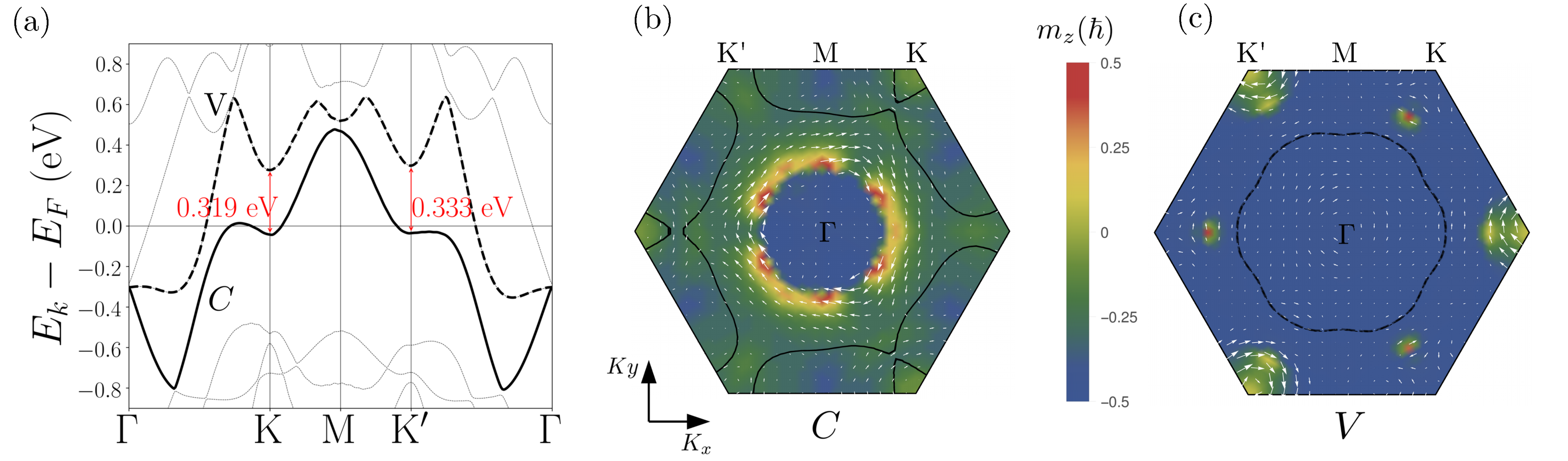}
\caption{(color online) Momentum dependent spin-polarization for the conduction and valence bands. (a) Fully relativistic band structure highlighting the two bands of interest. The thick solid line represents the conduction band labeled $C$ while the thick dashed line represent the valence band labeled V. The rest of the bands are represented with thin dashed lines with high transparency.(b) Momentum dependent spin polarization plotted in the full Brillouin zone. The background color represents the spin polarization along the z direction while the white arrows represent the in-plane components. The solid and dashed lines represent the calculated Fermi surfaces for $C$ and $V$, respectively.}
\label{fig:Spin-pol}

\end{figure*}

\section{Ground-state propreties.} Let us begin by analyzing Fig \ref{fig:Crystal-Structure}, which shows the crystal structure of the system consisting of a monolayer of WS$_{2}$ with an overlayer of Iron. Figs. \ref{fig:Crystal-Structure}(a) and \ref{fig:Crystal-Structure}(b) respectively represent the side and top views of the system, which consists of a stacking of 4 layers in a hexagonal honeycomb lattice structure where the Fe layer aligns itself on top of W. 
Our first principles calculations show that the system is ferromagnetic with a total magnetic moment of $2.02$ $\mu_{B}$, in agreement with previous works\cite{Polesya2016}. 
\

In Figure \ref{fig:electron-structure}, we present the calculated band structure (Fig. \ref{fig:electron-structure}(a)) and density of states (DOS) (Fig\ref{fig:electron-structure}.(b)) of Fe/WS$_2$. Let us first focus on the scalar relativistic band structure (i.e., without including SOC) represented by the dark (blue) and light (red) lines in Fig. \ref{fig:electron-structure}(a), which denote respectively spin-up and spin-down bands. As a first main feature, we note that only the spin-down bands cross the Fermi level, revealing that the system is half-metallic and thus attractive for spintronics \cite{HalfMet,Schmitt1987,Wolf2001,Fong2012,Jourdan2014}. This property can be further appreciated in Fig. \ref{fig:electron-structure}(b), where the calculated DOS reveals a gap of ~1 eV for the majority spin channel. As a second noteworthy feature of the scalar relativistic band structure, we note that the eigenenergies are symmetric under $\textbf{k}\rightarrow-\textbf{k}$ transformation, i.e $E_{n\bf k}=E_{n-\bf k}$.

 Next, we consider the effect of SOC in the band structure: this is illustrated by the dashed black lines of Fig. \ref{fig:electron-structure}(a), which correspond to a fully relativistic calculation. Two main effects can be remarked. The first one is a generalized and $\textbf{k}$-dependent shift of the bands with respect to the scalar relativistic calculation, reaching a maximum of $\sim120$ meV for the bands near the K and K$'$ points at around -0.7 eV (see Fig. \ref{fig:electron-structure}(a)). The second main feature revealed by the figure is that the eigenenergies are not degenerate at  $\textbf{k}$ and $-\textbf{k}$, i.e. $E_{n\bf k}\neq E_{n-\bf k}$, the so called valley-Zeeman splitting. This comes as a consequence of the combined effect of broken time reversal symmetry induced by the Fe overlayer and broken inversion symmetry characterized by a finite SOC\cite{Schaibley2016}. Noteworthily, this gives rise to a singular feature around the Fermi level, namely the appearance of a hole pocket near the K point (along $\Gamma-$K) that has no counterpart in the reversed  $\textbf{k}$-space region near K$'$ (along $\Gamma-$K$'$), where the corresponding band is totally occupied.

\
 We now turn to analyze the \textbf{k}-dependent spin-polarization, which is defined by
\begin{equation}\label{eq:K-dep.Mag}
\textbf{m}_n(\textbf{k})=\int \Psi^*_{n\textbf{k}}(\textbf{r})\textbf{$\hat{\sigma}$}\Psi_{n\textbf{k}}(\textbf{r}) d^3r,
\end{equation}
with $n$ the band index and $\Psi_{n \bf k}$ the spinor Bloch state. The above quantity is a convenient way of measuring the spin noncollinearity induced by SOC \cite{Eiguren2009,Ibanez-Azpiroz2011,Liu2009}.
We will focus our attention on the two specific bands represented in Fig. \ref{fig:Spin-pol}(a). These are the conduction ($C$) and valence ($V$) bands at K and K$'$ points, hence the most interesting ones concerning valley-related physics. The calculated spin-polarization for $C$ and $V$ is shown if Figs. \ref{fig:Spin-pol}(b) and \ref{fig:Spin-pol}(c) respectively, where arrows represent the in-plane components while the background describes the out-of-plane component. We also plot, on top of the spin-polarization, the Fermi surfaces corresponding to each band.
Noteworthily, the Fermi surface of $C$ (Fig. \ref{fig:Spin-pol}(b)) displays two cuts along $\Gamma-$K, which are induced by the aforementioned hole pocket present in the fully relativistic band structure (see Fig. \ref{fig:electron-structure}(a)). In contrast, there are no cuts of the Fermi surface along $\Gamma-$K$'$, revealing a clear asymmetry between the two regions.

Let us next note that, in absence of SOC, the spin-polarization of every state would be collinear. In particular, $C$ and $V$ would be polarized along the spin-down direction (see scalar relativistic bands in Fig. \ref{fig:electron-structure}(a)), implying that the illustration of  $\textbf{m}_{n}(\textbf{k})$  would be a totally blue map with no in-plane arrows. Therefore, any deviation from the blue background present in Figs. \ref{fig:Spin-pol}(b) and \ref{fig:Spin-pol}(c) must be attributed to the effect of SOC. Focusing first on $C$ (Fig. \ref{fig:Spin-pol}(b)), we note that the nearly circular bright spot structure encircling the $\Gamma$ point takes place at a band crossing, as it can be checked in Fig \ref{fig:Spin-pol}(a) along $\Gamma-$K and $\Gamma-$K$'$ at $\sim -0.8$ eV. The most interesting features for our purposes, however, are those close to the K and K$'$ points. Our calculations reveal that $C$ is spin-polarized approximately along the spin-down direction around both K and K$'$, with a roughly constant value of $\textbf{m}\simeq -0.1 \hbar \hat{\bf z}$. In contrast, $V$ (Fig. 3c) displays a totally different behavior; while the spin is almost completely polarized along the spin-down direction at K with $\textbf{m}\simeq -0.5 \hbar \hat{\bf z}$, the neighborhood of K$'$ displays a strong noncollinearity with substantial in-plane components. Exactly at the K$'$ point, however, the in-plane components vanish and the spin is polarized along the spin-up direction with $\textbf{m}\simeq +0.2 \hbar \hat{\bf z}$. This alternating spin-polarization structure at K and K$'$ points has a strong impact on light absorption, as we analyze in the next paragraphs. 

\begin{figure*}[t]
\centering
    
\includegraphics[width=0.8\textwidth]{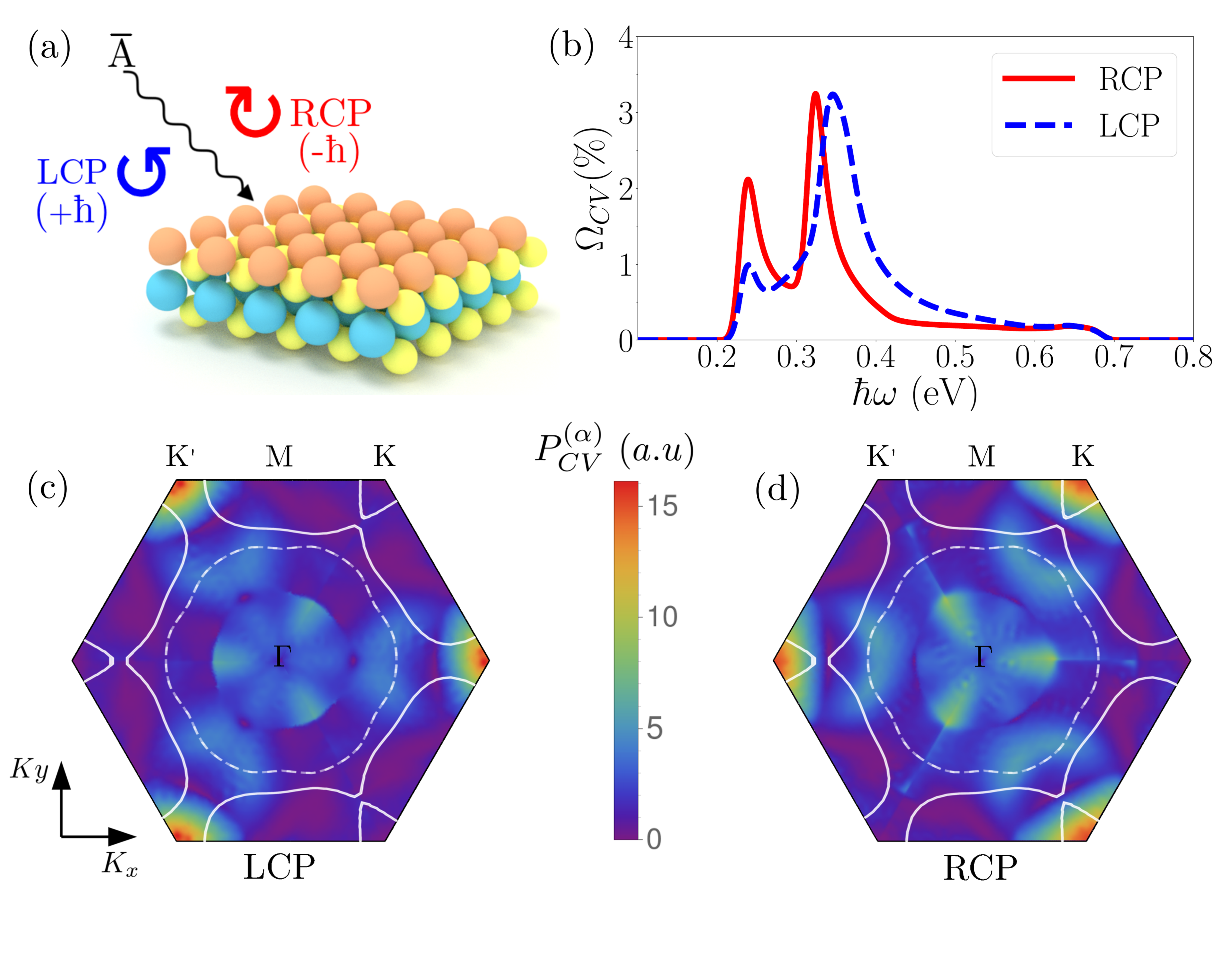}
\caption{(color online) (a) Schematic representation of left circularly polarized (LCP) and right circularly polarized (RCP) light excitation of Fe/WS$_{2}$, $\bar{A}$ represents the vector potential associated with the field. LCP and RCP carry spin angular momentum $+\hbar$ and $-\hbar$, respectively. (b) Calculated absorption rate associated with the spin-flip transitions, the blue dashed and red solid lines correspond to the absorption of left circularly polarized (LCP) and right circularly polarized (RCP) light, respectively. (a) and (b) Spin-flip transition probability associated  with the conduction band (C) and the valence band (V) for left circularly polarized (LCP) and right circularly polarized (RCP) light, respectively. The Fermi surface for each band is indicated by solid line for $C$ and a dashed line for V.}
\label{fig:MatrixElem}

\end{figure*}
\section{Transitions induced by circularly polarized light.} In this section we focus on the electronic transitions induced by an external time-dependent electric field and occur between the conduction and the valence bands, as well as the possibility of observing an associated magnetically dichroic signal in this material. We note that this type of response can be experimentally measured by polarization-resolved magneto-photoluminescence techniques \cite{Srivastava2015,Li2014}. Working within first order perturbation theory, we analyze the transition rate between $C$ and $V$ due to the absorption of a photon with frequency $\omega$ using Fermi's golden rule:

\begin{equation}\label{eq:tr-rate}
\begin{split}
\gamma^{(\alpha)}_{m  n  }(\omega)=&\dfrac{2\pi}{\hbar}\int 
\left( f(\epsilon_{{\bf k}m})-f(\epsilon_{{\bf k}n}) \right)
 |C^{(\alpha)}_{m n}(\textbf{k})|^{2} \\
&\times \delta(\epsilon_{{\bf k}n}-\epsilon_{\textbf{k}m}-\hbar\omega)\frac{d^{2}{\bf k}}{(2\pi)^{2}}
.
\end{split}
\end{equation}
Above, $f(E_{n\bf k})$ represents the Fermi-Dirac distribution function, while the integral is evaluated over the entire surface Brillouin zone. The transition matrix elements 
$T^{\alpha}_{CV}(k)$ are given by \cite{Rashba2003a,Ibanez-Azpiroz2013}

\begin{equation}
\label{eq:Matrix-Elems}
T^{(\alpha)}_{C V}(\textbf{k})=-\frac{e}{2c}\textbf{A}^{(\alpha)}_{0}\cdot\bra{\Psi_{\textbf{k} C}}\hat{\textbf{v}}
\ket{\Psi_{{\bf k}V}},
\end{equation}

with $\textbf{A}^{(\alpha)}_{0}$ the vector potential associated with the external electric field and $\alpha$ its polarization. In this work, we will focus on right circularly polarized (RCP) and left circularly polarized (LCP) light described by $\textbf{A}_{0}^{(R,L)}=A_{0}(\hat{\textbf{i}}\pm i\hat{\textbf{j}})/\sqrt{2}$, with A$_{0}$ the bare amplitude. We note that $\textbf{A}_{0}^{L}$ and $\textbf{A}_{0}^{R}$ carry spin angular momentum $+\hbar$ and $-\hbar$, respectively (see Fig. \ref{fig:MatrixElem}(a) for a schematic representation). 

The calculation of the matrix elements $T^{(\alpha)}_{mn} (\bf k)$ involves gradients in momentum space, which need a special treatment due to the inherent phase indeterminacy of the Bloch wave functions \cite{Blount1962,Wang2006}. In order to overcome this fundamental obstacle, we have made used of maximally localized Wannier functions \cite{Marzari1997,Souza2001,Mostofi2008} which ensures that the matrix elements entering Eq. (2) are smooth in \textbf{k} space. Following this procedure \cite{Ibanez-Azpiroz2012,Ibanez-Azpiroz2013}, we have, furthermore, been able to interpolate the  $T^{(\alpha)}_{CV}(\bf k)$ 
into a high density k-point grid needed to converge the integral of Eq. 
(2).

In Figs. \ref{fig:MatrixElem}(c) and \ref{fig:MatrixElem}(d) we present respectively the calculated transition probability for LCP and RCP light, defined as $ P^{(\alpha)}_{CV} (\textbf{k})\equiv |T_{CV  }^{(\alpha)}(\textbf{k})|^{2}/|\textbf{A}^{(\alpha)}_{0}|^{2}$  (see Eq. \ref{eq:Matrix-Elems}). 
The most important feature revealed by Figs. \ref{fig:MatrixElem}(c) and \ref{fig:MatrixElem}(d) is the high localization of the transition probability in hot-spot regions around the K$'$ and K points, respectively. Interestingly, the emergence of these hot-spots is closely related to the structure of the spin-polarization near the K and K$'$ points depicted in Figs. \ref{fig:Spin-pol}(b) and \ref{fig:Spin-pol}(c), as well as with the angular momentum transferred by the circularly polarized light. On one hand, LCP light transfers a total of $+\hbar$ angular momentum, hence the induced transitions are maximal in $\textbf{k}$-space regions where the spin-polarization is allowed to increase when excited from $C$ to $V$: the K$'$ point and its surroundings fulfill best this condition, given that the change in the spin-polarization there is  $\simeq +0.3\hbar $ . On the other hand, the reverse situation takes place for RCP polarized light in the neighborhood of the K point, where the spin-polarization decreases roughly by $\simeq -0.4\hbar$. Therefore, our calculations reveal that, when the effect of SOC is considered fully \textit{ab initio}, the transition probability strongly depends on the details of the non-collinear spin-polarization.

We next note that the magnitude of $P^{\alpha}_{CV}(\bf k)$ in the hot-spots of Figs. \ref{fig:MatrixElem}(c) and \ref{fig:MatrixElem}(d) is at least three times larger than in any other region of the Brillouin zone. Furthermore, the maximum probability in these regions is inequivalent for LCP (Fig. \ref{fig:MatrixElem}(c)) and RCP (Fig. \ref{fig:MatrixElem}(d)) light, being $\simeq 33\%$ larger in the case of the former. This turns out to be crucial for dichroic light absorption, given that the hot-spots lie in regions where $C$ is occupied while $V$ is unoccupied, as it can be clearly checked in Figs. \ref{fig:MatrixElem}(c) and \ref{fig:MatrixElem}(d) (see Fermi surface cuts). In other words, Fermi occupation factors effectively allow transitions  in regions with maximal probability, thus enhancing the magnitude of light absorption, which is analyzed in the next paragraph.

 \
 
A convenient way of measuring the light absorption associated to the incident light is given by the following expression,

\begin{equation}
\Omega^{(\alpha)}_{CV  }(\omega)= \frac{\hbar\omega\cdot\gamma^{(\alpha)}_{CV  }(\omega)}{\mathcal P }, 
\end{equation}
where ${\mathcal P}=|\textbf{A}^{(\alpha)}_{0}|^{2}\omega^{2}/8\pi c$ is the optical power per unit area of the incident field. $\Omega^{(\alpha)}_{CV  }(\omega)$ is therefore a measure of the percentage of light absorbed in excitation processes from $C$ to $V$.

The calculated absorption percentage $\Omega^{\alpha}_{CV}(\omega)$ is illustrated in Fig. \ref{fig:MatrixElem}(b) for LCP and RCP incident light. It is immediately apparent from the figure that light is absorbed inequivalently depending on the polarization and the energy of the incoming photon. Focusing first on the absorption peak located at $\sim 0.2$ eV, nearly 2$\%$ of the incoming light is absorbed when RCP light is shined, while the percentage decreases by half in the case of LCP, thus implying a strong dichroic signal at this energy. The origin of this peak can be attributed to a band-structure effect located midway along the $\Gamma-$K and $\Gamma-$K$'$ high symmetry lines, where $C$ and $V$ are quasi-parallel and separated by $\sim 0.2$ eV, giving rise to an interband peak at approximately this energy. At larger energies, Fig. \ref{fig:MatrixElem}(b) shows the maximum absorption peak of approximately $\sim 3.25\%$ at $\sim0.32$ eV for RCP  and at $\sim 0.34$ eV for LCP. As it is evidenced by Fig. \ref{fig:Spin-pol}(a), these energies correspond roughly to the band gaps at the K and K$'$ points: therefore, the contribution of the transition probability hot spots of Figs. \ref{eq:Matrix-Elems}(a) and \ref{eq:Matrix-Elems}(b) are the direct responsible for the maximum light absorption peaks of Fig. \ref{fig:MatrixElem}(b). 
Furthermore, we note that the full width at half maximum of the two peaks varies significantly, LCP being almost twice as big as RCP. This effect can once again be traced back to a band structure effect, namely the presence of a hole pocket for $C$ near high symmetry point K along $\Gamma-K$ (see Fig. \ref{fig:Spin-pol}(a)): this feature restricts the energy interval at which transitions take place and, moreover, has a predominant effect on the absorption of RCP light, as it is clearly visible from the Fermi surface cuts near the hot spots around K in Fig. \ref{fig:MatrixElem}(b). In contrast, the absence of such a hole pocket around K$'$ (see Figs. \ref{fig:Spin-pol}(a) and \ref{eq:Matrix-Elems}(a)) explains why LCP light is absorbed in a wider energy range, thus representing a neat example of how the combination of broken TR symmetry and strong SOC can affect to great extent the response properties of TMDs.
      
\

\section{Discussion and conclusions.} Our analysis has quantitatively demonstrated that the valley Zeeman effect, namely the inequivalent absorption of light by different valleys due to a magnetic shift,
can be achieved by coating TMDs with a magnetic overlayer such as Iron.
Noteworthily, this procedure would provide a stable platform for observing and further investigating this effect without the need of applying strong external magnetic fields as the ones recently employed in the experiments of Refs. \onlinecite{Aivazian2015} and \onlinecite{Srivastava2015}, while maintaining many of the interesting properties of TMDs. 
We note that a similar approach has been adopted in a very recent breakthrough experiment using a EuS substrate\cite{Zhao2017}.
In the particular system Fe/WS$_2$ analyzed in this work, our calculations have demonstrated a half-metallic behavior with a band gap of $~1$ eV for the spin majority channel, \textit{i.e.}, it maintains the semiconducting nature of WS$_2$ in one spin channel while it allows metallic charge carriers in the other one. Furthermore, the strong non-collinear spin-polarization induced by the large SOC associated to W and its effect on the dichroic light absorption revealed by our first principles analysis makes  Fe/WS$_2$ an outstanding candidate for further experimental and theoretical investigation
of the valley Zeeman effect and its manipulation via circularly polarized light.

\

\textit{Acknowledgements.}
We are very grateful to A. Eiguren for sharing with us a fully relativistic pseudopotential for W. This work has been supported by the Impuls und
Vernetzungsfonds der Helmholtz-Gemeinschaft Postdoc Programme, funding from the European Research Council (ERC) under the European Union's Horizon 2020 research and innovation 
programme (ERC-consolidator grant 681405 — DYNASORE) and funding from the Algerian ministry of higher education under the PNE programme. The authors gratefully acknowledge the computing 
time granted by the JARA-HPC Vergabegremium and provided 
on the JARA-HPC Partition part of the supercomputer JURECA at Forschungszentrum J\"{u}lich.

\bibliography{biblio}

\end{document}